\documentclass[pre,aps,twocolumn,amsmath,amssymb,amsfonts,floatfix,superscriptaddress]{revtex4-1} 
 
\usepackage{graphicx}
\usepackage{floatrow}
\usepackage{color}

\newcommand{\red}[1]{\textcolor{black}{#1}}

\def\ii{\mathbf{i}}
\def\j{\mathbf{j}}
\def\e{\varepsilon}
\def\s{s}

\definecolor{cream}{RGB}{222,217,201}

\begin{document}

\title{Experimental and Theoretical Bulk Phase Diagram and Interfacial Tension of Ouzo}

 \author{Andrew~J.~Archer}
\affiliation{Department of Mathematical Sciences and Interdisciplinary Centre for Mathematical Modelling, Loughborough University, Loughborough LE11 3TU, UK}
\author{Benjamin~D.~Goddard}%
\affiliation{School of Mathematics and the Maxwell Institute for Mathematical Sciences, University of Edinburgh, Edinburgh EH9 3FD, UK}%
\author{David~N.~Sibley}%
\affiliation{Department of Mathematical Sciences and Interdisciplinary Centre for Mathematical Modelling, Loughborough University, Loughborough LE11 3TU, UK}%
\author{James~T.~Rawlings}
\affiliation{Department of Physics and Mathematics, Nottingham Trent University, Clifton Lane, Nottingham NG11 8NS, UK}
\author{Ross~Broadhurst}
\affiliation{Department of Physics and Mathematics, Nottingham Trent University, Clifton Lane, Nottingham NG11 8NS, UK}
\author{Fouzia~F.~Ouali}
\affiliation{Department of Physics and Mathematics, Nottingham Trent University, Clifton Lane, Nottingham NG11 8NS, UK}
\author{David~J.~Fairhurst}
\email{david.fairhurst@ed.ac.uk}
\affiliation{School of Physics and Astronomy, University of Edinburgh, Edinburgh EH9 3FD, UK}

\begin{abstract}
Ouzo is a well-known drink in Mediterranean countries, with ingredients water, alcohol and trans-anethole oil. The oil is insoluble in water, but completely soluble in alcohol, so when water is added to the spirit, the available alcohol is depleted and the mixture exhibits spontaneous emulsification. This process is commonly known as the louche or Ouzo effect. Although the phase boundaries of this archetypal ternary mixture are well known, the properties of coexisting phases have not previously been studied. Here, we present a detailed experimental investigation into the phase behaviour, including tie-lines connecting coexisting phases, determination of the critical point (also called the plait point in ternary systems) and measurements of the surface tension and density for varying alcohol concentrations. Additionally, we present a theory for the thermodynamics and phase diagram of the system. With suitable selection of the interaction parameters, the theory captures nearly all features of the experimental work. This simple model can be used to determine both bulk and non-uniform (e.g.\ interfacial) properties, paving the way for a wide range of future applications of the model to ternary mixtures in general. We show how our accurate equilibrium phase diagram can be used to provide improved understanding of non-equilibrium phenomena.
\end{abstract} 

\maketitle

\section{Introduction}

The mixing of two immiscible liquids (e.g., water and oil) usually results in a phase separation, giving rise to two distinct fluid layers. However, it is often desirable to have an emulsion, consisting of a fine dispersion of droplets of one liquid within the other, often with bulk properties that can be quite distinct from the two phases individually. Due to their adjustable properties, emulsions find many industrial applications. Familiar edible emulsions include milk, salad dressing and mayonnaise \cite{mcclements2004food}, with many other emulsions used in the pharmaceutical \cite{khan2011basics} and oil industries \cite{kokal2005crude}, with perhaps the most well known example being emulsion paints. However, generating emulsions requires a large amount of energy in the formation of the interfaces between bulk and droplet, typically added to the system using high-speed mixers: Scholten et al.~\cite{scholten2008life} estimate that 1 joule of energy is required to produce a 100ml drink.

In very specific situations, a system will undergo {\it spontaneous emulsification}, immediately separating without the addition of external energy. 
Spontaneous emulsification was first reported by Johanne Gads in 1878~\cite{gad1878lehre} while investigating the reactions of bile salts: `regardless of any external shock, the mere mutual contact of certain liquids is sufficient to produce an emulsion which surpasses any milk in terms of fineness and uniformity' (translated from German). 
The most well-known modern demonstration is the Ouzo effect, named (somewhat arbitrarily) by Vitale and Katz \cite{Vitale2003} after a Greek alcoholic drink; many Mediterranean countries have their own version including Raki in Turkey, Pastis in France, Sambuca in Italy, Mastika in Bulgaria, Arak in Lebanon and notoriously Absinthe~\cite{lachenmeier2006absinthe} which could all have lent their name to the phenomenon.
In all versions, the pure spirit is a clear, single phase liquid comprising around 60\% water, 40\% ethanol (referred to henceforth as `alcohol') and a small amount of anethole, an aromatic ester (referred to as `oil'), which (along with other botanicals), gives the drink its distinctive aniseed taste. \red{Water and anethole are almost completely immiscible but both are fully soluble in alcohol.}
Moreover, given the choice between the two, alcohol has a much greater preference for water. 
In pure Ouzo, there is sufficient alcohol to solubilise the oil. 
However, when even a small amount of water is added, the alcohol partitions with the water, reducing the solubility of oil, leading to immediate precipitation of an oil rich phase. 
The spontaneous emulsification creates oil droplets of roughly a micron in size~\cite{Vitale2003, sitnikova2005spontaneously}.
A thorough review of spontaneous emulsification was published by Solans et al. in 2016 \cite{solans2016spontaneous}.
We highlight some of the key concepts below.
The thermodynamics of bulk phase separation is well understood, but why the droplets remain so long-lived is not fully understood \cite{roger2023nanoprecipitation}. 
The process is analogous to reducing the temperature of a water rich vapour leading to the formation of stable water droplets within clouds \cite{Vitale2003}.

\subsection{Ouzo Phase Behaviour}
Ouzo-related drinks have been around for at least three thousand years. They were mentioned in writings from ancient Egypt and Greece and featured in a 9th Century Roman poem. Ouzo was produced by 14th Century Greek monks, Raki was drunk in the Ottoman Empire in the 17th century, Absinthe was invented in Europe in the late 18th century, Sambuca was first distilled in Italy in 19th Century and Pastis in France in the early 20th Century.

Despite this rich history, the first scientific report into spontaneous emulsification was by Vitale and Katz \cite{Vitale2003} in 2003. 
Using divinyl benzene (DVB), an oil very similar to anethole, dissolved in water-alcohol mixtures they identified an `Ouzo region' in which a stable emulsion forms spontaneously.
This region was defined by samples with an oil to alcohol ratio greater than 1:20.
They plotted the phase diagram on logarithmic axes to highlight the very low oil concentration region.
The diameter of the spontaneously formed droplets ranged from 1 to 3 $\mu$m and were described by a logarithmic function of `excess' oil concentration above the 1:20 ratio, normalised by the alcohol concentration.
They attempted to make theoretical predictions, using two thermodynamic models, but noted that `unfortunately, lack of measured thermodynamic data on DVB, combined with the strong nonidealities of this three-component liquid system, made it impossible to calculate...'

In 2003, Grillo \cite{grillo2003small} performed neutron scattering measurements on anethole-water-alcohol emulsions, measuring the droplet diameter to be around 1$\mu$m immediately after formation, and increasing by 10\% after 12 hours with faster growth rates at higher temperatures.
Like Vitale and Katz, she also reported that the droplet size depends on the ratio of oil to alcohol.

In 2005 Sitnikova and co-workers \cite{sitnikova2005spontaneously} also investigated the phase behaviour and droplet stability of the anethole-water-alcohol system, producing a detailed phase diagram exploring the stable Ouzo region. 
They used dynamic light scattering to track droplet diameters, finding a slow increase from 1$\mu$m to 3$\mu$m over 16 hours.
The droplet volume increased linearly over time, indicating growth via diffusion of dissolved oil across the continuous phase (Ostwaldt ripening) rather than through droplet coalescence (which exhibits exponential growth).
Interestingly, they observed the droplets undergoing Brownian motion, and frequently colliding without merging.
The measured growth rates allowed the authors to estimate the interfacial tension of the emulsion droplets to be between 0.1 and 100 mN~m$^{-1}$, depending strongly on composition.
We compare their deduced surface tension values with our direct measurements below in Fig.~\ref{fig:st_compare}.
Following the growth stage (around 60 hours), the emulsion was stable for several months, showing a very slow reduction in droplet size, attributed to sedimentation or creaming of the droplets.
They ascribed the long-term stability of these emulsions to the low interfacial tension but note that the behaviour is strongly dependent on the mixing process, claiming that adding water to a single phase system is the only way to create a stable emulsion. 
They also suggest that `the presence of small amounts of a residual charge on the droplet may in principle prevent the droplets from merging'.
They also propose that ethanol has a stabilising effect on the emulsion by accumulating at the interface, much like a surfactant.

In 2009 Scholten et al.~\cite{scholten2008life} performed stability experiments using anethole-water-ethanol and commercial Pernod.
They identified the impossibility of directly measuring the interfacial tension between oil droplets and arbitrary water/ethanol mixtures due to the constraints of coexistence.
They mitigated this problem to some extent by pre-saturating oil droplets in the mixtures before performing pendant drop measurements, but by doing so, sacrificed knowledge of the exact compositions of both phases.
They report interfacial tension values of 11 mN~m$^{-1}$ for 30\% ethanol samples and 1.4 mN~m$^{-1}$ for 70\% ethanol samples and use these values to estimate the ripening rate (the reverse of the calculation in Ref.~\cite{sitnikova2005spontaneously}).
Using their measurements for various ternary mixtures, they predict sedimentation/creaming rates to be between 5 and 35 nm~s$^{-1}$. 
However, their incomplete knowledge of coexisting phases limits further discussions: in particular, they incorrectly invoke critical behaviour to explain their observations, without identifying the location of the critical point in the phase diagram.

All previous works have reduced ethanol concentration to trigger spontaneous emulsification by dilution with water; however the same result can also be achieved by allowing ethanol to evaporate.
Tan et.al~\cite{tan2016evaporation} explore this effect in micro-litre droplets.
They report compositional data for the cloud-point curve where precipitation first occurs.
We compare their data to ours in Fig.~\ref{fig:binodal}.
In his 2018 PhD thesis~\cite{tan2018evaporation} Tan calculates the compositions of coexisting phases in the Ouzo system and plots them as tie-lines on the phase diagram. 
Tan uses this to predict the location of the critical point to be at weight fractions of ethanol $w_e = 35\%$, water $w_w =8\%$ and oil $w_o=57\%$.
He uses diffusion path theory (see below) to predict the time evolution of dilution and evaporation.

On the theoretical front, in 1972, Ruschak and Miller produced the landmark paper explaining spontaneous emulsification \cite{ruschak1972spontaneous}.
They consider generic phase diagrams and highlight how `diffusion paths' linking the two phases that are brought into contact can be used to predict emulsion formation, either with or without bulk phase separation.
Provided the diffusion path crosses through the two phase region, producing a region of supersaturation, then an emulsion will form.
They use three toluene-water-solute ternary systems to illustrate their predictions.
Krishna \cite{krishna2015serpentine} extends the theoretical framework work with `serpentine' trajectories, by allowing for concentration-dependent diffusion coefficients.
However, this additional consideration is not necessary to explain the Ouzo effect.

For over 40 years the formation of transparent (non-scattering) micro-emulsions with dimensions around 2nm, in a region close to the Ouzo region has been a curious mystery.
However in 2016 Zemb et al.~\cite{ zemb2016explain} proposed a model for the behaviour in this `pre-Ouzo region' and
using neutron-scattering data from the ethanol-water-octanol ternary system, and molecular dynamics simulations they demonstrate how small clusters of octanol are stabilised by a coating of ethanol.
From their model, they argue that this can be predicted by considering effects of hydration and entropy.

Very recently, Roger et al.~\cite{roger2023nanoprecipitation} show how the mixing method affects the size of emulsion droplets and that particularly rapid mixing can remove the Ouzo region altogether.
They consider a ternary hexadecane-acetone-water system, but their findings should be universal.
They use Raman spectroscopy to measure the compositions of coexisting samples, allowing experimentally-determined tielines to be plotted.

\subsection{Calculating Phase Diagrams}
A prerequisite to a full understanding and to having a model for long-lived emulsions is access to a complete description of the equilibrium thermodynamics of the system, as a function of the concentrations of the mixture components. 
Then, building on this, one can subsequently work to understand the non-equilibrium thermodynamics of emulsions or any other non-equilibrium aspects of Ouzo-type systems -- e.g.\ the beautiful and striking dynamics that arise when water droplets are deposited in oil-alcohol mixtures \cite{DFD_movie}.
The equilibrium thermodynamics of any system is determined by the free energy (and its derivatives), from which  one can obtain all relevant measurable thermodynamic quantities and, perhaps most importantly, the bulk phase diagram of the system. 
The approach taken here is to perform experiments in order to accurately determine the bulk phase diagram, and in parallel to develop a simple model for the free energy. 
We confirm the accuracy of our model by comparing predictions for the phase diagram with the experimental observations. 
Our model applies not just to the bulk fluid phase behaviour; it can also be used to determine features of the inhomogeneous fluid -- i.e.\ to determine interfacial properties, such as the surface tension or wetting behaviour. 
Our model consists of a simple lattice density functional theory (DFT) for the ternary mixture. 
A derivation of the model for a one component system is given in Hughes et al.~\cite{hughes2014introduction}, which also explains the Picard iteration method that we use to solve the equations to obtain the fluid density profiles. 
The lattice DFT that we use may be viewed as being based on classical DFT for continuum fluids \cite{evans1979nature, hansen2013theory} and as being a coarse-grained (discretized) approximation to a continuum DFT. 
Lattice DFT has been used to describe one-component liquids on surfaces \cite{hughes2015liquid, buller2017nudged}, adsorption in porous media \cite{kierlik2001capillary, woo2001mean, schneider2014filling}, binary liquid mixtures on surfaces \cite{woywod2003phase, WoSc2004jpm, robbins2011modelling, archer2010dynamical, chalmers2017modelling, areshi2024binding, perez2021changing}, colloid aggregation behaviour \cite{edison2015critical, edison2015phase, tasios2016critical} and to determine the stability of nanoparticle laden aerosol droplets \cite{archer2023stability}.
Here, we extend the model to describe ternary mixtures, with the interaction parameters chosen carefully to mimic the Ouzo system.
As mentioned, we validate the model by demonstrating that it correctly captures the bulk phase diagram and we also use it to calculate the surface tension of the free interface and the densities of the coexisting phases.
With this, as future work, the model can then be applied to describe the behaviour of the ternary mixture in contact with planar surfaces (c.f.\ Ref.~\cite{areshi2024binding}) and also in non-equilibrium studies by using the free energy we develop here together with dynamical DFT \cite{chalmers2017dynamical, hansen2013theory, te2020classical}.

\subsection{Phase Behaviour Overview}

In our experiments, the natural quantity for characterising the composition of a particular sample is the mass fraction. This is due to the fact that it is straightforward to measure the total mass of each of the components added to the mixture, $M_{\rm a}$, $M_{\rm o}$ and $M_{\rm w}$, the masses of alcohol, oil and water, respectively. The total mass of liquid is then of course $M=M_{\rm a}+M_{\rm o}+M_{\rm w}$. Thus, the mass fraction of component $p$ is $w_p=M_p/M$, where $p$ is one of a, o, w.

In contrast, in our lattice DFT introduced below in Sec.~\ref{sec:4}, the natural quantities are dimensionless {\em number} densities of each species in the system, defined as $n^p=N_p/V$, where $N_p$ is the number of molecules of species $p$ in the system and $V$ is a dimensionless total volume. Note that the reason that we put the species label $p$ as a superscript is that the local density can be a spatially varying quantity: In our lattice DFT, we denote position in our system via a discrete lattice vector $\ii =(i,j,k)$, where $i$, $j$ and $k$ are integers. With this, the spatially varying local density distribution of species $p$ is denoted as $n_\ii^p$. Note also that in our dimensionless units, the volume $V$ is also the total number of lattice sites in the system, where we use the lattice spacing $\sigma$ as the length scale (unit of length) in our model.

Closely related to these are the {\em mass} densities of the three species, defined as $\rho_{\rm a}=n^{\rm a}m_{\rm a}$, $\rho_{\rm o}=n^{\rm o}m_{\rm o}$ and $\rho_{\rm w}=n^{\rm w}m_{\rm w}$, respectively, where $m_{\rm a}$, $m_{\rm o}$ and $m_{\rm w}$ are the corresponding masses per unit volume of the three liquids. The total mass density is then of course $\rho=\rho_{\rm a}+\rho_{\rm o}+\rho_{\rm w}$. In order to compare with the experiments, we convert from our dimensionless number densities $n^p$ to mass densities using the following values for the densities of the three pure liquids: $m_{\rm a}=0.79$ g/cm$^3$, $m_{\rm o}=0.99$ g/cm$^3$ and $m_{\rm w}=1.00$ g/cm$^3$. Converting in this way does not take into account the fact that when you mix two liquids, e.g.\ alcohol and water, the final volume is not the sum of the two separate volumes. But, given all the other approximations in our model (see Sec.~\ref{sec:4}), for present purposes converting in this way is satisfactory.

In any mixture, the mass fractions $w_p$ are not independent, but must add to 100\%. The usual way to plot the phase diagram of three-component systems such as the Ouzo mixture of interest here is to use the following coordinate transform to map the individual mass fractions onto a triangular ternary phase diagram:
\begin{align}
x=\frac{1}{2}\frac{2w_{\rm o}+w_{\rm a}}{w_{\rm w}+w_{\rm a}+w_{\rm o}}, \nonumber \\
y=\frac{\sqrt{3}}{2}\frac{w_{\rm a}}{w_{\rm w}+w_{\rm a}+w_{\rm o}}.
\label{eq:1}
\end{align}
Note that the above formulae also apply if the mass fractions are replaced by the mass densities (i.e.\ multiply both top and bottom of the above by $M/V$).

\begin{figure*}[!t]
\centering
\includegraphics[width=1.00\linewidth,keepaspectratio]{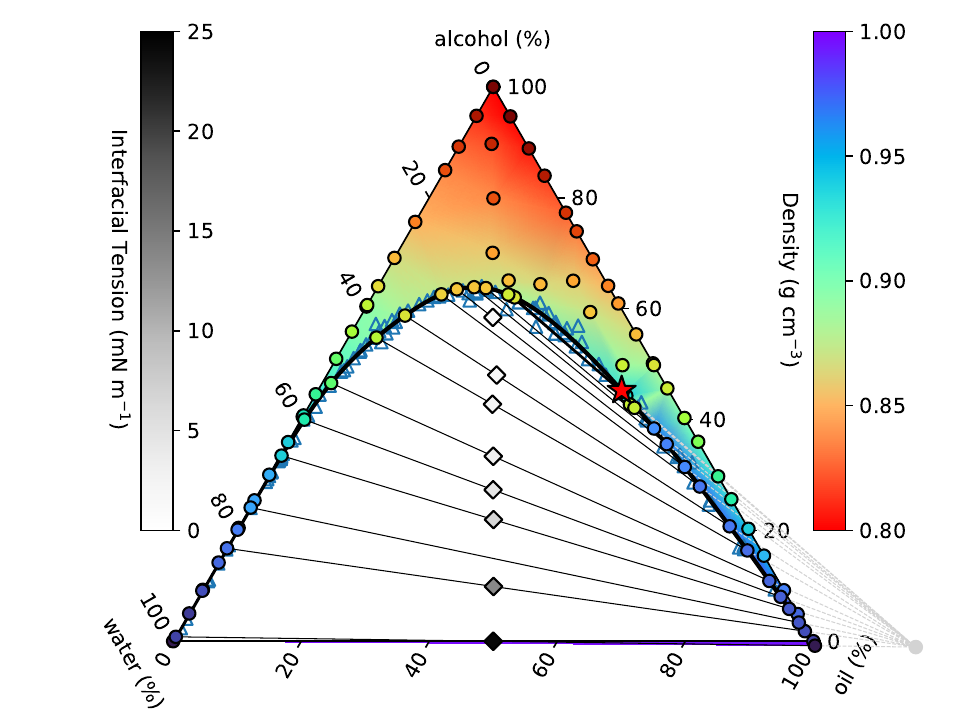}
\caption{Measured Ouzo phase diagram. The thick black curve is the binodal given by the fit in Eq.~\eqref{eqn:binodal}. Individual density measurements are shown by coloured circles, and the coloured background shading is a smooth extrapolation between the points. The thin black lines are tie-lines between coexisting states. These are generated using the origin identified from the fit discussed in Sec.~\ref{sec:3.3} (see also Figs.~\ref{fig:binodal} and \ref{fig:tielines}) and indicated by the light grey circle and dashed lines in the lower right. The varying slope of the tie-lines clearly highlights the alcohol preferentially partitioning into the water-rich phase. Diamonds indicate the initial samples prepared for density and surface tension measurements, and are shaded according to the measured surface tension. The critical point is identified with a red star.}
\label{fig:AllTernary}
\end{figure*}

In Fig.~\ref{fig:AllTernary} we present our measured ternary phase diagram for the water, alcohol and oil mixture and use this figure to introduce the key features of the phase diagram. We postpone discussion of the experimental details of the materials used and the approach used to obtain this diagram till later. The benefit of this triangular representation of a ternary mixture phase diagram is that every point on it shows the overall composition of the mixture, with the three corners representing the three pure phases.\cite{davis1996statistical} Here, alcohol is at the top,  pure water on the left and pure oil on the right. 

The solid black line in Fig.~\ref{fig:AllTernary} is the binodal curve, separating stable mixtures (above) from those that phase-separate (below). \red{The binodal is also referred to by some as the cloud curve, as samples just below the line often become cloudy due to precipitation of the second phase. Of course, other phase separation phenomena also occur in this region of the phase diagram, such as spinodal decomposition leading to macroscopic phase separation}. For our Ouzo system, essentially all of the horizontal water-oil axis lies beneath the binodal, indicating that water and oil are completely immiscible. Conversely, both left and right hand sides of the triangle are above the binodal, since alcohol is completely miscible in both water and oil. 

Samples inside the binodal separate into two distinct phases, with defined compositions located on the binodal. This equilibrium does not depend on the amount of either phase, so by varying the relative fractions of each of the coexisting phases, we are able to trace out the tie-line of all initial compositions that phase-separate to give the two end-points of the specific tie-line. The so-called {\it lever rule} states that the amount of each coexisting phase depends on the distance between the point representing the overall composition and the two end-points. An initial sample with composition very close to one end separates primarily into the phase at the near endpoint, with only a small amount of the further endpoint phase. Initial samples in the middle of a tie-line separate into equal amounts (mass or volume depending on the representation used) of the two phases.

On many (but not all) ternary phase diagrams, there is a critical point, or plait point, at which the composition of the two phases becomes identical \cite{davis1996statistical}. Consequently, on approaching the critical point, the tie-lines become very short. Our experimentally-determined critical point is denoted in Fig.~\ref{fig:AllTernary} by a red star.

Immediately within the binodal curve, samples are metastable, and can only fully phase separate after nucleation of a droplet of the minority phase occurs.
This nucleation process has an associated free energy barrier to surmount, whereby a droplet of the minority phase larger than a certain critical nucleus size, must occur \cite{debenedetti1996metastable}. Once the minority phase is nucleated, it can grow without restriction. However, further into the two phase region, beyond the spinodal boundary, the sample becomes thermodynamically unstable to even the smallest fluctuation in density and the sample immediately phase separates.

In a two-component system in which temperature controls the phase behaviour, the vertical axis in a phase diagram is naturally temperature, with horizontal tie-lines illustrating thermal equilibrium. In ternary systems, the tie-lines represent equal chemical potentials between the coexisting phases. In our case, alcohol chemical potential (or equivalently alcohol concentration) plays an analogous role to temperature. However, it is not possible to measure chemical potentials directly. The tie-lines are slanted, indicating partitioning of alcohol between the water-rich and oil-rich phases. The composition of the coexisting phases can be determined through repeated chemical analysis of samples extracted from each phase. We implement a simpler method using knowledge of the position of the binodal along with the lever rule \cite{capela2019simple}.

\subsection{Overview}
The remainder of this paper is structured as follows: In Sec.~\ref{sec:2} we give the details of the experimental methodology used to determine the Ouzo phase diagram in Fig.~\ref{fig:AllTernary}. Then, in Sec.~\ref{sec:3} we discuss the experimental results obtained and our methodology for fitting the data to simple expressions for the binodal curve and also the tie-lines between coexisting phases. We also present our results for the surface tension, comparing with the results from our simple lattice DFT. Following this, in Sec.~\ref{sec:4}, we give a detailed description of our lattice DFT, including how to calculate the bulk fluid phase diagram, showing comparison with the experimental results. We also briefly explain how to calculate the inhomogeneous fluid density profiles and so obtain the interfacial tension. 
Finally, in Sec.~\ref{sec:5}, we show how improved knowledge of the equilibrium behaviour allows us to make predictions about the dynamics of the Ouzo behaviour, followed by concluding remarks.

\section{Experimental Methodology}
\label{sec:2}

\subsection{Materials}

Ultra-pure water, of resistivity 18.2 M$\Omega$, produced using a Purelab Chorus I water purifier, along with HPLC gradient grade ethanol (purity $\ge$99.8\%), supplied by Fluka Analytical and trans-anethole oil (purity $\ge$99\%), purchased from Sigma Aldrich, were used in this study. All chemicals were used as supplied unless otherwise stated.

\subsection{Binodal curve}

The oil-rich side of the binodal curve was first investigated using the standard `cloud curve' method with the results displayed in Figs.~\ref{fig:binodal} and \ref{fig:tielines}. A single phase oil-alcohol solution was made (in the oil-rich side of the phase diagram) and diluted incrementally with measured quantities of water until the system was observed to phase separate, typically into a cloudy emulsion. The composition of this phase-separating `cloud point' sample was recorded as the location of the binodal. The emulsion was then diluted with known amounts of alcohol and oil to return to a different location in the single-phase region. It was again diluted with water to form a new emulsion and the process repeated following a zig-zag path around the binodal. A similar process was performed to investigate the water-rich portion of the phase diagram, this time starting with a water-alcohol mixture and adding oil.

\subsection{Coexisting Samples}

To determine the properties of the coexisting samples, we prepared over twenty solutions with equal masses of water and oil, and varying amounts of alcohol, within the two-phase region. The samples were thoroughly mixed before being left to separate. They were then centrifuged for two minutes to ensure complete separation of the two phases. A pipette was used to carefully extract all of the upper and lower phases, and the mass of each phase, $M_1$ and $M_2$, was measured. In the pure water-oil system, with no alcohol, water is the more dense (lower) phase. However, once only a modest amount of alcohol is added, due to alcohol's preferred solubility in water, the water-rich phase becomes less dense. This cross-over of densities can be seen where the curves cross in Fig.~\ref{fig:density}. For samples very close to the critical point the interface between phases remained diffuse. Extraction of the upper and lower phases was performed very carefully, but it was not possible to measure the interfacial tension or the masses of the two phases.

\subsubsection{Density}

The mass density $\rho$ of each of the coexisting phases was measured using an Anton Paar DMA4500 density analyser, which provides readings to 4 decimal places, and typically very good repeatability to 3 decimal places. The instrument was calibrated using pure water and pure ethanol before use and flushed clean using ethanol and dry air between each sample.

\subsubsection{Interfacial Tension}
We used a Kr{\"u}ss DSA100B Drop Shape Analyser to determine the interfacial tension between coexisting samples. The extracted lower phase was loaded into a capillary and used to form a pendant droplet in a reservoir of the extracted upper phase. Measurement of the gravitationally-deformed interface, along with previously determined values of the densities of the two phases allows for an accurate calculation of the interfacial tension. This method was carried out for a range of initial alcohol concentrations.

\section{Results}
\label{sec:3}

\subsection{Binodal Curve}

\begin{figure}[t!]
\centering
\includegraphics[width=1\linewidth,keepaspectratio]{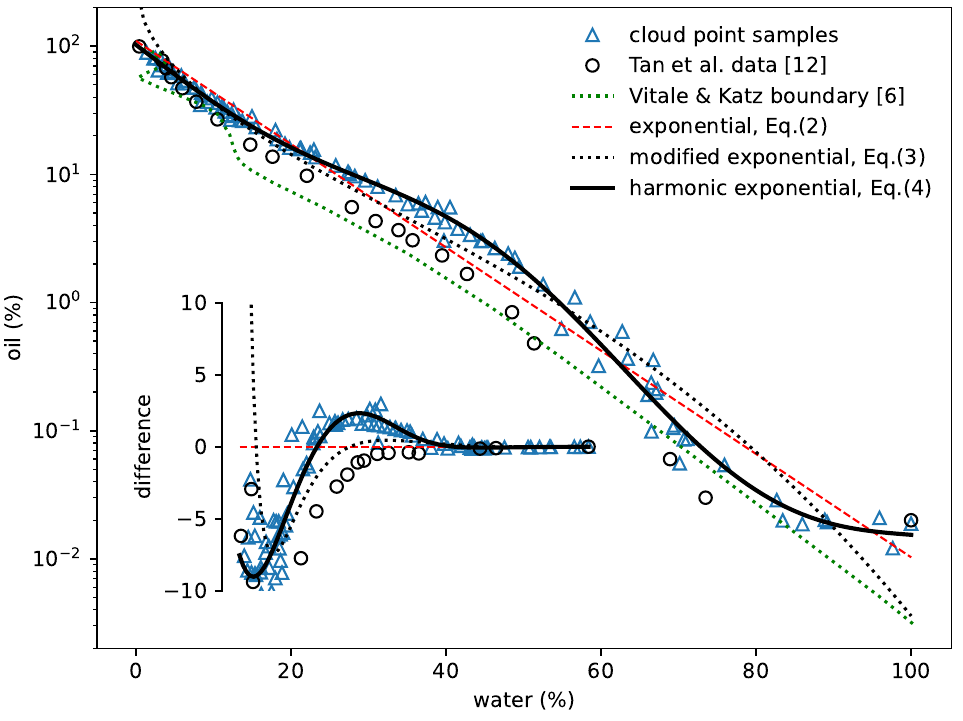}
\caption{Experimentally measured cloud point samples (blue triangles) together with data taken from Tan et al.~\cite{tan2016evaporation} (black circles) and the boundary provided by Vitale and Katz \cite{Vitale2003} Fits to the three expressions in Eqs.~\eqref{eqn:exp}, \eqref{eqn:sqrtcubic} and \eqref{eqn:binodal} are also shown. The inset shows the same data points and curves subtracted from the simple exponential fit in Eq.~\eqref{eqn:exp}.}
\label{fig:binodal}
\end{figure}

In total around 100 cloud point samples were identified to determine the location of the binodal.
The compositions of these samples are  plotted in Fig.~\ref{fig:binodal} and values provided in the Appendix, in Table \ref{tab:binodal}. 
Our measurements are in good agreement with the experimental work of Tan et~al.~\cite{tan2016evaporation}. 

We find that the following simple exponential analytic expression can be used to fit the data for the mass fraction of oil $w_{\rm o}$ as a function of water mass fraction $w_{\rm w}$ along the binodal
\begin{equation}
    w_{\rm o} = a\exp{\left(b w_{\rm w}\right)}.
    \label{eqn:exp}
\end{equation}
The fit parameters $a$ and $b$ depend on the regression method used, with average values $a=0.97\pm0.05$, $b=-8.8\pm0.6$ and $R^2=0.98$. Others \cite{capela2019simple,freire2012aqueous,merchuk1998aqueous} regularly use an alternative of the form 
\begin{equation}
  w_{\rm o} = a\exp{\left(bw_{\rm w}^{0.5} - c w_{\rm w}^3\right)}
  \label{eqn:sqrtcubic}
\end{equation}
to fit binodal curves. However, this gives a worse fit to our data. Instead, we find an improved fit by including the first three terms of a Fourier sine series:
\begin{equation}
w_{\rm o} = a\exp{\left(bw_{\rm w} + c \sin{\pi w_{\rm w}} + d \sin{2\pi w_{\rm w}} + e \sin{3\pi w_{\rm w}}\right)},
\label{eqn:binodal}
\end{equation}
with $a=1.03\pm0.06$, $b=-8.80\pm0.12$, $c=12\pm9$, $d=46\pm5$ and $e=-47\pm4$. The alcohol concentration for any sample can be calculated from the result $w_{\rm a} = 1 - w_{\rm w} - w_{\rm o}$.

\subsection{Tie-lines}
To determine the coexisting compositions, we follow the method of Ref.~\cite{capela2019simple} and combine our empirical expression for the binodal in Eq.~\eqref{eqn:binodal}, with the lever rule which balances the masses of coexisting phases ($M_1$ and $M_2$) based on distance in phase space from the two phase boundaries.

\begin{figure*}[t!]
\centering
\includegraphics[width=1\textwidth,keepaspectratio]{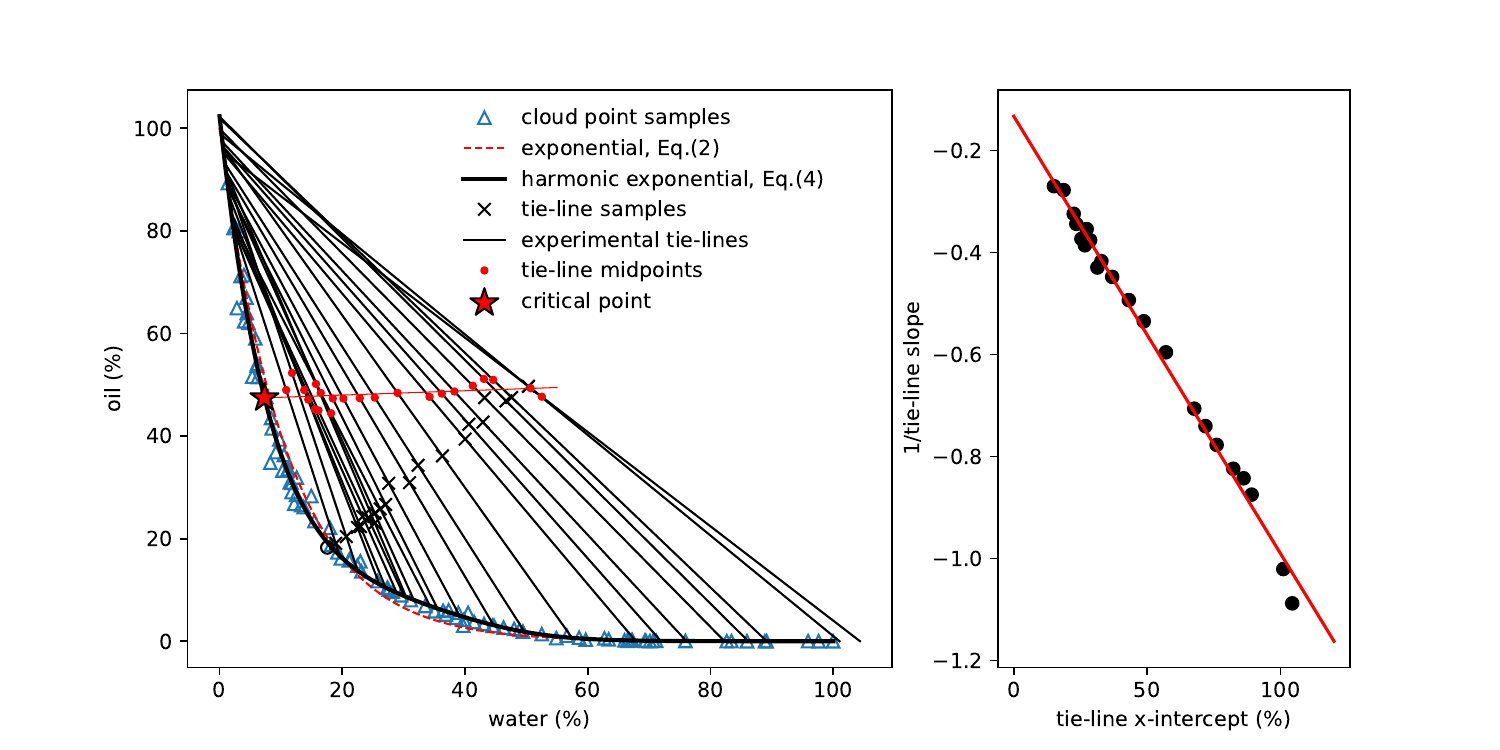}
\caption{Determining experimental tie-lines. Left: The blue triangles show the cloud-point samples (as in Fig \ref{fig:binodal}), the dashed red line is the exponential fit~\eqref{eqn:exp} and the thick black line the fit using Eq.~\eqref{eqn:binodal}. The black crosses show initial samples used for tie-line determination, which are connected to the two coexisting phases by thin black tie-lines. The hollow circle is the one sample for which coexisting phases could not be determined. Red dots indicate the midpoints of each tie-line and the thin red line a linear fit to these points. This line crosses the binodal at the critical point. The properties of these coexisting samples are presented in Table \ref{tab:properties} Right: A plot of the reciprocal slope of the (black) tie-lines against their extrapolated intercept on the horizontal $w_{\rm w}$ axis. The slope and intercept of the red linear fit surprisingly allows determination of a single point which acts as the origin for the tie-lines.}
\label{fig:tielines}
\end{figure*}

For samples close to the binodal, the lever rule predicts more of the `nearby' phase and less of the `distant' phase, in a ratio given by the reciprocal of the distances to the boundaries. Mathematically this can be written as
\begin{equation}
M_1 (w_{\rm 1w}-w_{\rm iw}) = M_2 (w_{\rm 2w}-w_{\rm iw})
\label{eqn:lever1}
\end{equation}
where the subscripts $i$, 1 and 2 refer to the initial sample and the two coexisting phases. The other constraint of the lever rule is that the two sections of the tie-line must have the same slope:
\begin{equation}
\frac{w_{\rm 1w}-w_{\rm iw}}{w_{\rm 1wo}-w_{\rm io}} = \frac{w_{\rm 2w}-w_{\rm iw}}{w_{\rm 2wo}-w_{\rm io}}.
\label{eqn:lever2}
\end{equation}
These two equations have four unknowns, the water and oil concentrations in each of the two coexisting phases. We then use Eq.~\eqref{eqn:binodal} to write the oil concentrations as a function of the water concentration, reducing the number of unknown quantities to two. The equations are then solved numerically.

Figure \ref{fig:tielines} shows the initial samples and the calculated properties of the two coexisting phases. The tie-lines all have gradient greater than 1, indicating that alcohol is preferentially dissolved in the water phase; a gradient of 1 corresponds to equal alcohol in both phases. A few of the tie-lines can be seen to cross each other: physically, this is not allowed and is due to uncertainties in the experimental determination of the phase boundary, sample preparation, physical separation and mass measurement of each phase. There are also some endpoints \red{(e.g.\ the bottom right point on Fig.~\ref{fig:AllTernary} and the top two tielines in Fig.~\ref{fig:tielines})} which have concentrations of water slightly below 0\% or above 100\%, which is of course also unphysical. Finally, for one sample, indicated by a hollow circle, it was not possible to mathematically find a solution to the coexistence equations. Despite these minor inconsistencies reflecting experimental uncertainties, the tie-lines show a clear picture of alcohol partitioning.

We indicate the midpoint of each calculated tie-line with a red dot on Fig.~\ref{fig:tielines}. The critical point is identified by fitting these points with a straight line and finding where it intersects the binodal. We find the critical point to have composition of $w_{\rm w}=7.33\%$, $w_{\rm o}=47.42\%$ and $w_{\rm a}=45.25\%$. For comparison, Tan \cite{tan2018evaporation} predicted, $w_w =8\%$  oil $w_o=57\%$, $w_a = 35\%$, which agrees with our water value but has more oil and less alcohol than our measurement.

Visual inspection of the tie-lines suggests that they are all emanating from a single point in the upper left hand region of the plot. To test whether this is a meaningful observation, we plot in Fig.~\ref{fig:tielines} the reciprocal of the slope of each tie-line against where, if extrapolated, it would intercept on the x-axis. The slope and intercept of this good straight line fit allow us to calculate the `origin' point of all the tie-lines to have composition of $w_{\rm w}=-15.49\%$, $w_{\rm o}=116.56\%$ and $w_{\rm a}=-1.07\%$. This purely empirical observation of an `origin' point does not carry any physical significance, but is a useful method to predict the slope of tie-lines. \red{The line from this origin is tangent to the binodal at the critical point, as expected.} It would be interesting to see if other ternary systems have a similar behaviour. We note that our lattice DFT calculations presented later do not find such an origin point.

\subsection{Properties of Coexisting Phases}
\label{sec:3.3}

Our method for determining the tie-lines required manually separating the upper and lower phases. This allowed for further measurements of the properties of the coexisting phases, in particular their densities and the surface tension between them.

\subsubsection{Density}

The measured density values for the coexisting phases are plotted in Fig.~\ref{fig:density} and the values provided in Table \ref{tab:properties} in the Appendix. The x-axis is somewhat arbitrary, and is one way of characterising the coexisting phases we are considering. The benefit of using the mass fraction difference $(w_{\rm w}^1-w_{\rm w}^2)$ is that coexisting samples have the same x-value coordinate and so allows for easier comparison. The mass fraction difference is also related to the alcohol chemical potential, which is also equal in coexisting phases.

\begin{figure}[b!]
\centering
\includegraphics[width=1.00\linewidth,keepaspectratio]{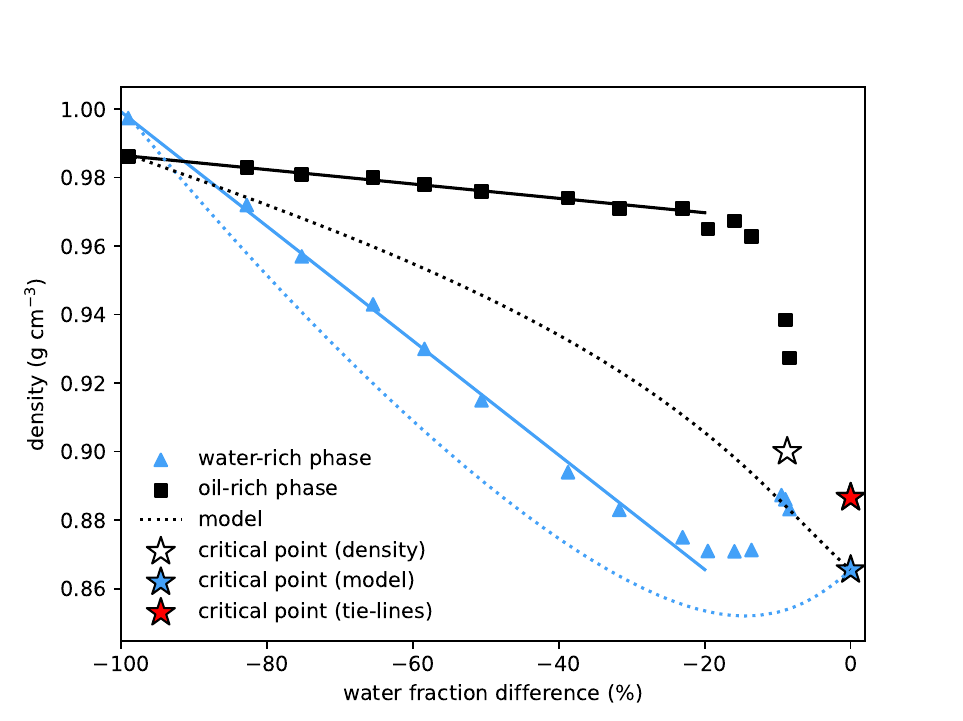}
\caption{Density of coexisting phases, as a function of the water mass fraction difference $(w_{\rm w}^1-w_{\rm w}^2)$ between the two phases. The density difference between the oil-rich and water-rich phases initially increases roughly linearly (fits added), but becomes zero at the critical point, the location of which we estimate by extrapolating the density data (white star). The water-rich data compares well with density data of water-alcohol mixtures taken from Ref.~\cite{scholten2008life}. The dotted lines show the predictions from the model presented in Sec.~\ref{sec:4}, with the blue star indicating the critical point from the model. These show good qualitative agreement, with better agreement for the water-rich phase. The critical point determined from the tie-line midpoints, with density determined by extrapolating between nearby measurements, is plotted as a red star.}
\label{fig:density}
\end{figure}

In agreement with literature values, \red{for the binary system with no added alcohol, we measure the almost pure water sample to have a slightly higher density than the almost pure coexisting oil sample: on the figure, the left-most blue triangle is above the coexisting black square, both at water mass fraction difference equal to $-100$\%}. Of particular interest is the observation that, owing to the preferential interaction between water and alcohol (which has a significantly lower density), the water rich phase is less dense for all other coexisting samples. Interpolating the data, we estimate that coexisting phases have the same density at around 3\% initial alcohol concentration.

The density difference continues to increase with alcohol concentration (increasing water mass fraction difference) as the water preferentially dissolves a larger fraction of the alcohol. However, we know that this increasing density difference cannot continue indefinitely and must decrease again to zero on reaching the critical point. The final data points at water mass fraction differences of around 10\% indeed show this.
Verifying this prediction experimentally is difficult due to the expected \cite{ hansen2013theory, kawasaki1970kinetic, bray2002theory} slow phase separation dynamics close to a critical point: both the surface tension and the density difference go to zero so the driving force for macroscopic phase separation is reduced. Without a sharp interface, it is experimentally difficult to cleanly extract the two coexisting phases.
We extrapolate the experimental data points to find another estimate for the critical point indicated by the white star. 
For comparison, we add a blue star for the model critical point and a red star for the critical point from Fig.\ref{fig:AllTernary}, with density value calculated by interpolating between the three closest experimental measurements. 
Our values compare qualitatively well with density values for binary and ternary mixtures provided in Ref.~\onlinecite{scholten2008life}.
A quantitative comparison is not possible as the exact compositions of their coexisting samples are not determined.

\subsubsection{Interfacial Tension}

\begin{figure}[t!]
\centering
\includegraphics[width=1.00\linewidth,keepaspectratio]{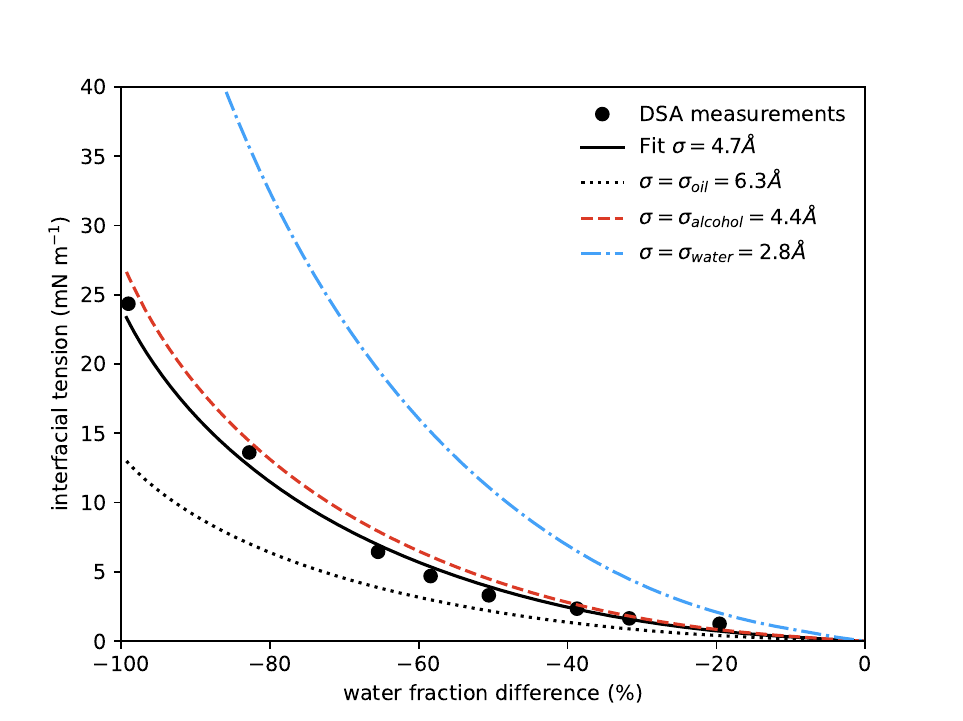}
\caption{Interfacial tension measured between coexisting phases using drop shape analysis as a function of the water mass fraction difference between the two coexisting phases. The symbols are the experimental results. The solid black line shows the prediction from our model at a temperature of $20^\circ$C, using only one fitting parameter, the size of the lattice sites $\sigma$, which is found to be 0.47nm. The additional curves show the predicted interfacial tension using instead for $\sigma$ the literature values for the size of molecules of water (blue dash-dots, 0.275nm), alcohol (red dash, 0.44nm) and anethole oil (black dots, 0.63nm). Similar surface tension data was presented in Ref.~\cite{scholten2008life}, showing qualitatively similar behaviour, but it is not possible to determine appropriate values to plot on our axes and compare directly.}
\label{fig:st}
\end{figure}

Our measurements of the interfacial tension for the interface between coexisting samples was made using a Kr\"{u}ss DSA100B Drop Shape Analyser. These are plotted in Fig.~\ref{fig:st} as a function of the water mass fraction difference $(w_{\rm w}^1-w_{\rm w}^2)$ between the two phases and also provided in the Appendix, in Table \ref{tab:properties}. We see that the surface tension decreases as the water mass fraction difference decreases (or, equivalently, with increasing alcohol concentration), and, as expected, approaches zero as the critical point is approached. Our measurements \red{show the same qualitative trend as the results} reported in Ref.~\cite{scholten2008life}.

In Fig.~\ref{fig:st} we also present results for the surface tension calculated using the lattice model described in the following section. The model has a single free parameter $\sigma$, the lattice spacing.
A diagram of the set-up of the model is shown in Fig.~\ref{fig:diagram}, which illustrates how $\sigma$ is most naturally interpreted as the typical size of the space occupied by one of the molecules in the mixture. Choosing $\sigma=0.47$nm, we find (see Fig.~\ref{fig:st}) that our lattice model is able to very accurately match the experimental results. Although this value of $\sigma$ is chosen by treating it somewhat arbitrarily as fit parameter that can be chosen to give best agreement with the experiments, it is gratifying to see that this value is physically appropriate. Recall the literature values for the size of water molecules (0.28nm), alcohol (0.44nm) and oil (0.63nm). Thus, our best-fit value falls right in the middle of the physically-reasonable range. Indeed, using any one of these three other values instead still produces good agreement, as shown in Fig.~\ref{fig:st}.

Our surface tension results are also presented in Fig.~\ref{fig:AllTernary}, where we shade the diamond symbols on the tie-lines according to the measured surface tension. This again illustrates the decrease in the surface tension on approaching the critical point. Having now presented all our experimental results for the Ouzo system phase behaviour and also comparing in Figs.~\ref{fig:density} and \ref{fig:st} for the coexisting density values and surface tension obtained from our lattice DFT, we now move on to present the model used to obtain these results.

\section{Lattice DFT model}
\label{sec:4}

\begin{figure}[t!]
   \centering
   \includegraphics[width=0.8\linewidth]{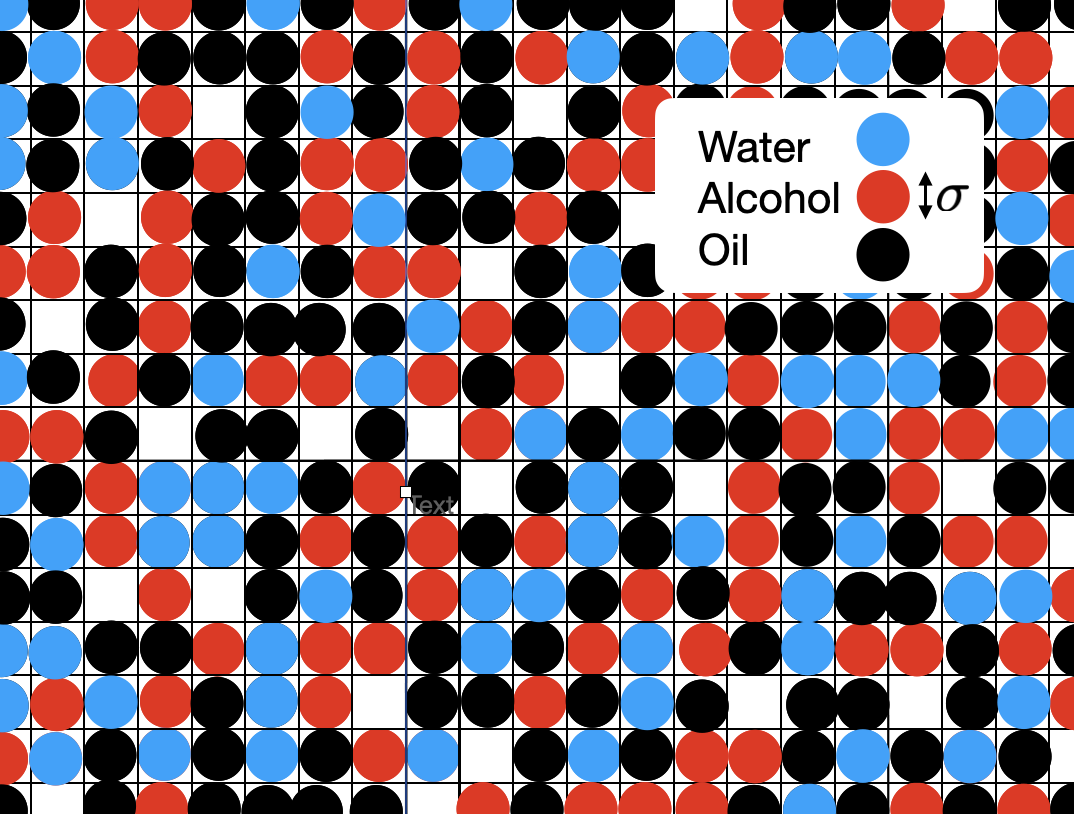}
   \caption{
   To model a three component mixture of water, alcohol and oil, we coarse-grain the system onto a lattice, as illustrated. We choose the size of each lattice site $\sigma$ to correspond roughly to the diameter of the alcohol molecules. Lattice sites are described as either occupied with water (blue circle), occupied by alcohol (red circle), occupied by oil (black circle) or empty (white).}
   \label{fig:diagram}
\end{figure}

In Fig.~\ref{fig:diagram} is an illustration of the molecules in a water-alcohol-oil mixture, i.e.\ in Ouzo. To model the system, we discretize onto a lattice, choosing the lattice spacing $\sigma$ to be the diameter of the space typically occupied by one of the molecules in the mixture. Of course, the different molecules are different in size, so $\sigma$ should be thought of as some kind of `average' size. We then assume that each lattice site either contains either oil, water, alcohol or is empty. By `contains water', we really mean `contains mostly water'. Following Ref.~\cite{chalmers2017dynamical}, this discretization enables us to map the system onto a three-species lattice-gas model (generalised Ising model) which then enables to investigate the thermodynamics of the system. See also the related kinetic Monte-Carlo model discussed in Ref.~\cite{botet2012ouzo}.

We denote the location of each lattice site by the index $\ii$. In three dimensions, we have $\ii =(i,j,k)$, where $i$, $j$ and $k$ are integers. We introduce three occupation numbers for each lattice site, $\ell_\ii^{\rm w}$, $\ell_\ii^{\rm a}$ and $\ell_\ii^{\rm o}$, for the water, alcohol and oil, respectively. If a lattice site is empty, then all three $\ell_\ii^{\rm w}=\ell_\ii^{\rm a}=\ell_\ii^{\rm o}=0$. If lattice site $\ii$ is occupied by water, then $\ell_\ii^{\rm w}=1$, while $\ell_\ii^{\rm a}=\ell_\ii^{\rm o}=0$. If instead lattice site $\ii$ contains alcohol, then $\ell_\ii^{\rm a}=1$ and $\ell_\ii^{\rm w}=\ell_\ii^{\rm o}=0$. The final possibility is that site $\ii$ contains oil, then $\ell_\ii^{\rm o}=1$ and $\ell_\ii^{\rm w}=\ell_\ii^{\rm a}=0$. We assume it is impossible for $\ell_\ii^{\rm w}+\ell_\ii^{\rm a}+\ell_\ii^{\rm o}>1$, i.e.\ any given lattice site can only contain at any given moment one of the three species. The potential energy of the system can then be expressed as
\begin{align}
\label{eq:Hamiltonian}
E =& - \sum_{\ii,\j}\left(
\frac{1}{2} \e_{\ii\j}^{\rm ww} \ell_{\ii}^{\rm w} \ell_{\j}^{\rm w}
+\frac{1}{2} \e_{\ii\j}^{\rm aa} \ell_{\ii}^{\rm a} \ell_{\j}^{\rm a}
+\frac{1}{2} \e_{\ii\j}^{\rm oo} \ell_{\ii}^{\rm o} \ell_{\j}^{\rm o}\right. \nonumber \\
& \qquad \qquad \left.+ \e_{\ii\j}^{\rm wa} \ell_\ii^{\rm w} \ell_\j^{\rm a}
+ \e_{\ii\j}^{\rm wo} \ell_\ii^{\rm w} \ell_\j^{\rm o}
+ \e_{\ii\j}^{\rm ao} \ell_\ii^{\rm a} \ell_\j^{\rm o}
\right) 
\nonumber
\\
&+ \sum_\ii \left(\Phi_\ii^{\rm w} \,\ell_\ii^{\rm w} + \Phi_\ii^{\rm a} \ell_\ii^{\rm a}+ \Phi_\ii^{\rm o} \ell_\ii^{\rm o}\right),
\end{align}
which is a sum over interactions between neighbouring lattice sites. The six matrices $\e^{pq}$, where $\{p,q\}\in \{\rm w,a,o\}$, define the pair interactions between different lattice sites.\cite{chalmers2017modelling} These are essentially discretized pair potentials. Note that with the minus sign on the pair interaction terms, we are following the convention that a matrix $\e^{pq}$ with positive values corresponds to {\em attractive} interactions between particles. The terms in the final line of Eq.~\eqref{eq:Hamiltonian} are contributions due to any external potentials $\Phi_\ii^p$ acting on the three different species, e.g.\ due to any container walls. Here these are assumed to be zero, but for future application of the model these may be non-zero.

\subsection{Free Energy}
We follow Refs.~\cite{hughes2014introduction, hughes2015liquid, chalmers2017dynamical} to develop a theory for the ensemble-averaged densities
\begin{align}
n_{\ii}^{\rm w} = \langle \ell_\ii^{\rm w}\rangle,\,\,\,\,\,\,\,n_\ii^{\rm a} = \langle \ell_\ii^{\rm a}\rangle \,\,\,\,\,\,\,\text{and} \,\,\,\,n_\ii^{\rm o} = \langle \ell_\ii^{\rm o}\rangle,
\end{align}
giving the following approximation for the Helmholtz free energy
\begin{align}
F
  = & k_B T \sum_\ii \Big[
        n^{\rm w}_\ii \ln n^{\rm w}_\ii
    + n^{\rm a}_\ii \ln n^{\rm a}_\ii
    + n^{\rm o}_\ii \ln n^{\rm o}_\ii
    \nonumber \\
    & \qquad \qquad + (1 - n^{\rm w}_\ii - n^{\rm a}_\ii-n^{\rm o}_\ii) \ln (1 - n^{\rm w}_\ii - n^{\rm a}_\ii-n^{\rm o}_\ii)\Big]
  \nonumber\\
  &- \sum_{\ii,\j}\Big(
\frac{1}{2} \e_{\ii\j}^{\rm ww} n_{\ii}^{\rm w} n_{\j}^{\rm w}
+\frac{1}{2} \e_{\ii\j}^{\rm aa} n_{\ii}^{\rm a} n_{\j}^{\rm a}
+\frac{1}{2} \e_{\ii\j}^{\rm oo} n_{\ii}^{\rm o} n_{\j}^{\rm o} \nonumber\\
&\qquad \qquad
+ \e_{\ii\j}^{\rm wa} n_\ii^{\rm w} n_\j^{\rm a}
+ \e_{\ii\j}^{\rm wo} n_\ii^{\rm w} n_\j^{\rm o}
+ \e_{\ii\j}^{\rm ao} n_\ii^{\rm a} n_\j^{\rm o}
\Big)
\nonumber
\\
&\quad
+ \sum_\ii \left(\Phi_\ii^{\rm w} \,n_\ii^{\rm w} + \Phi_\ii^{\rm a} n_\ii^{\rm a}+ \Phi_\ii^{\rm o} n_\ii^{\rm o}\right),
\label{eq:helmholtz}
\end{align}
where $k_B$ is Boltzmann's constant and $T$ is the temperature. For non-zero $\Phi_\ii^p$ this free energy may be used to calculate inhomogeneous fluid density profiles. However, to begin with, we focus our interest on the bulk fluid phase behaviour, where the densities are uniform, i.e.\ $n_\ii^{\rm w}=n^{\rm w}$, $n_\ii^{\rm a}=n^{\rm a}$ and $n_\ii^{\rm o}=n^{\rm o}$, are constants for all $\ii$. From Eq.~\eqref{eq:helmholtz} we find that the Helmholtz free energy per unit volume, $f = F/V$, where $V$ is the volume of the system, is given by:
\begin{align}
\label{eq:f_bulk}
f = & k_BT\big[n^{\rm w} \ln{n^{\rm w}}+n^{\rm a} \ln{n^{\rm a}}+n^{\rm o} \ln{n^{\rm o}}\nonumber\\
& \qquad + (1 - n^{\rm w} - n^{\rm a} - n^{\rm o}) \ln(1 - n^{\rm w} - n^{\rm a} - n^{\rm o})\big]
\nonumber
\\ 
& - \frac{1}{2}\s^{\rm ww}(n^{\rm w})^2
- \frac{1}{2}\s^{\rm aa}(n^{\rm a})^2
- \frac{1}{2}\s^{\rm oo}(n^{\rm o})^2\nonumber\\
&- \s^{\rm wa} n^{\rm w} n^{\rm a}
- \s^{\rm wo} n^{\rm w} n^{\rm o}
- \s^{\rm ao} n^{\rm a} n^{\rm o},
\end{align}
where $\s^{pq}$ is the integrated strength (i.e.\ sum over all entries) of the pair potential matrix $\e^{pq}$, i.e.\ $\s^{pq}=\sum_{\j}\e_{\ii\j}^{pq}$. The bulk fluid phase behaviour is therefore determined by the temperature and the values of the six parameters $\s^{\rm ww}$, $\s^{\rm aa}$, $\s^{\rm oo}$, $\s^{\rm wa}$, $\s^{\rm wo}$ and $\s^{\rm ao}$. Depending on the values of these, there can be complete mixing of all species, gas-liquid phase separation and also liquid-liquid phase separation. For the case of Ouzo, the oil and the water do not like to mix, so we expect $\s^{\rm wo}$ to be the smallest in value of all, but the alcohol can mix with either the water or with the oil, so we expect the remaining $\s^{pq}$ to all be somewhat comparable in value. A mixture of all three can often also demix, but this depends on the concentrations of the three species. There is also a vapour phase.

For our discretized pair potential $\e_{\ii\j}^{pq}$, we follow Refs.~\cite{chalmers2017modelling, chalmers2017dynamical}, by setting it to have the form $\varepsilon_{\mathbf{ij}}^{pq} = \epsilon_{pq} c_{\mathbf{ij}}$, where the interaction tensor between species-$p$ and species-$q$ particles has the form
\begin{equation}
\label{eq:c_ij_liquids}
c_{\mathbf{ij}}  = 
  \begin{cases} 
   1 & \text{if }\mathbf{j}\in {NN \mathbf{i}}, \\
    \frac{3}{10}    & \text{if }\mathbf{j}\in {NNN \mathbf{i}}, \\
    \frac{1}{20}    & \text{if }\mathbf{j}\in {NNNN \mathbf{i}}, \\
        0   & \text{otherwise},
         \end{cases}
\end{equation}
where $NN\mathbf{i}$, $NNN\mathbf{i}$ and $NNNN\mathbf{i}$ denote the nearest neighbours of $\mathbf{i}$, next nearest neighbours of $\mathbf{i}$ and next-next nearest neighbours of $\mathbf{i}$, respectively. The six parameters $\e^{pq}$, where $\{p,q\}\in \{\rm w,a,o\}$, determine the overall strength of the pair potentials. Thus, the values of these potential strength parameters are simply related to the integrated strengths of the pair potentials, via $\s^{pq}=10\e^{pq}$. The particular choices in Eq.~(\ref{eq:c_ij_liquids}) mean that the liquid-liquid or liquid-vapour surface tension of the model has only a small dependence on the orientation of the interface with respect to the underlying lattice \cite{kumar2004isotropic, chalmers2017modelling, chalmers2017dynamical}. In other words, when we use the model to calculate the shape of liquid droplets on surfaces, the droplet interfaces are almost perfectly hemispherical, as expected \cite{archer2023stability, areshi2024binding}.

\subsection{Bulk Fluid Phase Behaviour}

Returning to considering the bulk fluid phase behaviour, we recall that for two phases to coexist, the temperature $T$, pressure $P$ and chemical potentials $\mu^p$ of all three species $p$ must be equal in the two coexisting phases,\cite{hansen2013theory} which we denote here as phase 1 and phase 2. In other words, the five conditions: 
$T_1=T_2$, $P_1=P_2$ and $\mu_1^p=\mu_2^p$, for $p\in\{{\rm a,o,w}\}$, must be satisfied.
The phase diagram consists of a plot of the coexisting states. These form a line (or surface in a 3D representation) in the phase diagram, referred to as the binodal. The three chemical potentials are obtained as
\begin{equation}
\mu^p = \frac{\partial f}{\partial n^p},
\label{eq:mu}
\end{equation}
while the pressure (or equivalently minus the grand potential density) is obtained from the relation
\begin{equation}
P = -f+\mu^{\rm w}n^{\rm w}+\mu^{\rm a}n^{\rm a}+\mu^{\rm o}n^{\rm a}
\label{eq:p}.
\end{equation}
To satisfy the five coexistence conditions, the first ($T_1=T_2$) is trivially easy to satisfy, since $T$ is just a parameter in Eq.~\eqref{eq:f_bulk}. To satisfy the remaining four conditions, we then need to find three density values $\{n^{\rm w}_1,n^{\rm a}_1,n^{\rm o}_1\}$ in phase 1 and simultaneously the densities $\{n^{\rm w}_2,n^{\rm a}_2,n^{\rm o}_2\}$ in phase 2. In other words, this at first appears to be trying to solve four equations for six unknowns. However, two of the conditions on the chemical potentials can be split, giving the following six equations
\begin{align}
P_1=P_2, \hspace{1cm}
\mu_1^{\rm w}=\mu_2^{\rm w}, \nonumber\\
\mu_1^{\rm a}=\hat{\mu}^{\rm a}, \hspace{1cm}
\mu_2^{\rm a}=\hat{\mu}^{\rm a}, \nonumber\\
\mu_1^{\rm o}=\hat{\mu}^{\rm o}, \hspace{1cm}
\mu_2^{\rm o}=\hat{\mu}^{\rm o},
\label{eq:coex_conditions3}
\end{align}
where the chemical potentials of the alcohol $\hat{\mu}^{\rm a}$ and the oil $\hat{\mu}^{\rm o}$ are specified a priori. These six equations [together with Eqs.~\eqref{eq:f_bulk}, \eqref{eq:mu} and \eqref{eq:p}] can then be solved for the six densities $\{n^{\rm w}_1,n^{\rm a}_1,n^{\rm o}_1,n^{\rm w}_2,n^{\rm a}_2,n^{\rm o}_2\}$. We do this using the fsolve function in Maple. There is only a solution for a finite range of values of $\hat{\mu}^{\rm a}$ and $\hat{\mu}^{\rm o}$. Once a solution for one set of parameters is found, it can then be tracked as one of the parameters is varied. The easiest way to start this is to consider the limit where $n^{\rm a}\to0$ (i.e.\ just an oil-water mixture) that demixes at sufficiently low temperatures (or equivalently sufficiently large value of $\e^{\rm wo}$), where $\hat{\mu}^{\rm a}\to-\infty$. Setting $\beta\hat{\mu}^{\rm a}=-10$ is sufficient, where $\beta=(k_BT)^{-1}$.
The coexisting water and oil densities (essentially pure phases, with only a small value for the other species) together with a very small initial guess values for $n^{\rm a}_1$ and $n^{\rm a}_2$ obtains the solution in this limit. Then, on increasing $\hat{\mu}^{\rm a}$ for fixed $\hat{\mu}^{\rm o}$, the full binodal line for that temperature and value of $\hat{\mu}^{\rm o}$ can be mapped out. The full binodal surface can then be mapped out by repeating the above process for a series of values of $\hat{\mu}^{\rm o}$.

In experiments on Ouzo, the vapour phase is often not of interest and the liquid phase behaviour is the quantity of interest. In this case, the above approach can be simplified somewhat by making the assumption that the liquid is incompressible and that the particle number densities satisfy the condition
\begin{equation}
n^{\rm w} + n^{\rm a} + n^{\rm o} =1.
\end{equation}
We then introduce two concentration variables
\begin{align}
\phi=n^{\rm w} - n^{\rm a} + n^{\rm o},\nonumber \\
\eta=n^{\rm w} + n^{\rm a} - n^{\rm o},
\end{align}
which of course is equivalent to setting $n^{\rm w}=(\phi+\eta)/2$, $n^{\rm a}=(1-\eta)/2$ and $n^{\rm o}=(1-\phi)/2$. Substituting these into Eq.~\eqref{eq:f_bulk}, we can then obtain the coexistence conditions as equality of the `chemical potentials'
\begin{align}
\mu^\phi = \frac{\partial f}{\partial \phi},\nonumber \\
\mu^\eta = \frac{\partial f}{\partial \eta},
\label{eq:mu_phi_eta}
\end{align}
in the two phases. Similarly, the pressure
\begin{equation}
P = -f+\mu^\phi\phi+\mu^\eta\eta
\label{eq:pressure}
\end{equation}
is equal in the two phases. Exactly as described above, we solve by writing the coexistence conditions as
\begin{align}
P_1=P_2, \hspace{1cm}
\mu_1^\phi=\mu_2^\phi, \nonumber\\
\mu_1^\eta=\hat{\mu}^\eta, \hspace{1cm}
\mu_2^\eta=\hat{\mu}^\eta,
\label{eq:coex_conditions2}
\end{align}
i.e.\ by splitting the last condition and specifying the starting value of the chemical potential $\hat{\mu}^\eta$. The limit where there is no alcohol in the system is again a good place to start, i.e.\ in the limit $\eta\to1$. Then, once the coexisting values in that limit is obtained, then the value of $\hat{\mu}^\eta$ can be varied, in order to follow the binodal to the critical point.

For the pair potential parameter values used here (given below), the phase diagram calculated for the compressible system [i.e.\ by solving Eqs.~\eqref{eq:coex_conditions3}] is almost indistinguishable from the phase diagram calculated for the incompressible model [i.e.\ solving Eqs.~\eqref{eq:coex_conditions2}] over a broad range of values of the oil chemical potential, specifically for $-5<\beta\mu_{\rm o}<2$. Thus, we solely present here the results for the phase diagram from the incompressible model. However, for our surface tension calculations, we use the compressible model and set $\beta\mu_{\rm o}=-2$, which is a value roughly in the middle of the range where the differences between the phase diagrams are almost invisible.

The experimental phase diagram of Ouzo in Fig.~\ref{fig:AllTernary} is somewhat unusual and the main challenge in selecting appropriate values of the six interaction strength parameters $\e^{pq}$, in order for the theory to best match the experiments, is to correctly locate the critical point, far down on the right hand side of the binodal. It is trivial to select parameters which lead to the critical point being around the maximum on the binodal curve (point on the binodal where the alcohol concentration is highest), but to find it located on one side is much more difficult to achieve within our model. Another feature of the experimental phase diagram is the fact that the gap between the edge of the phase diagram and the binodal is much smaller on the left hand side than on the right hand side. This feature is discussed further below. With much trial and error and also drawing experience from previous work \cite{chalmers2017modelling, hughes2014introduction, hughes2015liquid, chalmers2017dynamical, areshi2019kinetic, areshi2024binding} we find the following values give surprisingly good agreement with the experimental phase diagram:
\begin{align}
&\beta\e^{\rm ww}=0.96,\hspace{0.5cm} &\beta\e^{\rm wa}=0.84 \nonumber\\
&\beta\e^{\rm aa}=0.78,\hspace{0.5cm}
&\beta\e^{\rm ao}=0.63 \nonumber \\
&\beta\e^{\rm oo}=0.78,\hspace{0.5cm}&\beta\e^{\rm wo}=0.30.
\label{eq:epsilons}
\end{align}

\begin{figure}[t!] 
   \centering
   \includegraphics[width=1.0\linewidth]{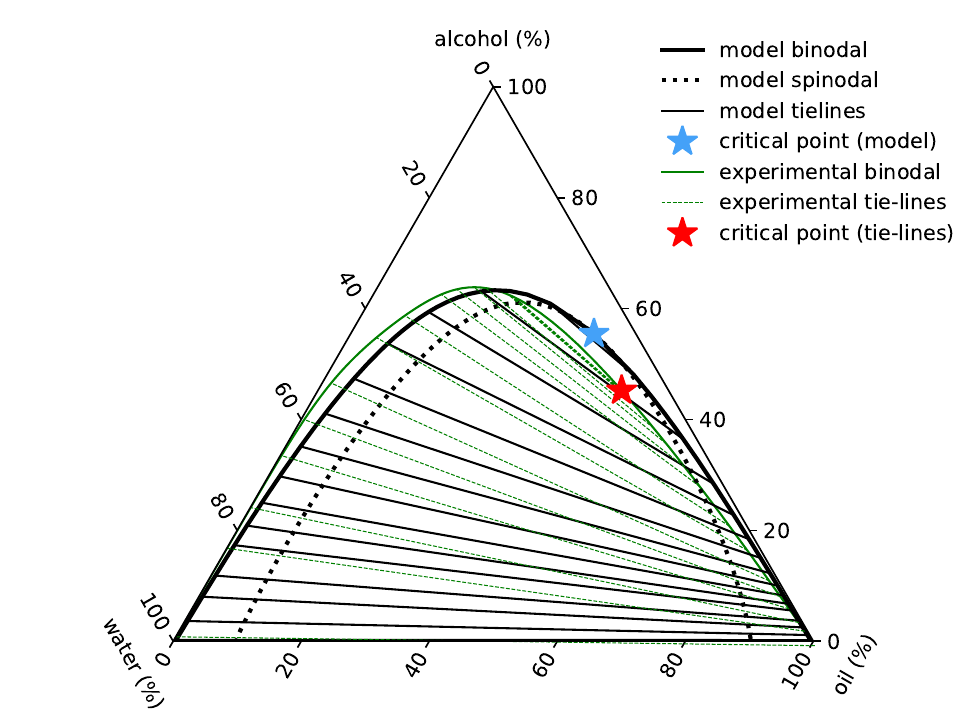} 
   \caption{Ouzo phase diagram with $\e_3=0$. Black lines indicate the results of the model binodal (solid) spinodal (dotted), tie-lines (thin) and critical point (blue star). The experimental results are in green (with the full set of results displayed in Fig.~\ref{fig:AllTernary}).}
   \label{fig:phase_diag}
\end{figure}

The resulting phase diagram is plotted in Fig.~\ref{fig:phase_diag}, showing the comparison with the experimental binodal. To understand the values selected for the $\e^{pq}$, first consider the water-oil mixture. Oil and water phase separate, therefore the energy for water to be adjacent to oil must be much less than the water-water or oil-oil energy. Thus, we choose $\e^{\rm wo}$ to be a small value (less than half the value of $\e^{\rm ww}$ and $\e^{\rm oo}$). Since water has hydrogen bonding, we expect $\e^{\rm ww}>\e^{\rm oo}$, hence the choice above. We choose for simplicity $\e^{\rm aa}=\e^{\rm oo}$. Water and alcohol happily mix, therefore a value $\e^{\rm wa}\gtrsim(\e^{\rm ww}+\e^{\rm aa})/2$ is appropriate \cite{hansen2013theory}. Finally, we varied the value of the parameter $\e^{\rm ao}$ to locate the critical point in roughly the right place, compared to the experimental phase diagram. Of course, application of an optimisation algorithm to fit these parameters would surely give an even better agreement. However, we found the learning experience of doing it by hand to be valuable!

The phase diagram displayed in Fig.~\ref{fig:phase_diag} is plotted in the same way as Fig.~\ref{fig:AllTernary}, using the coordinate transforms in Eq.~\eqref{eq:1}, so that each of the three pure liquids correspond to the three corners, while the distance from a given corner yields the local mass density of that species. Recall also our discussion just above Eq.~\eqref{eq:1} relating to how we convert from the number densities in our model to mass densities that can be compared with the experimental results.

The main feature in the experimental phase diagram that our simple model fails to capture is the distance between the left hand binodal and the edge of the diagram, i.e.\ when there is plenty of water present, our model over-predicts the amount of oil in the system. We believe this is related to the strength and directionality of the hydrogen bonding in water. Details like this are not present in our simple model. One of the signatures of hydrogen bonding is the tetrahedral ordering of the water molecules. In other words, three-body interactions are important \cite{hansen2013theory}. We have assumed just pair interactions. To test this hypothesis, we ad-hoc added a term $-\e_3(n^{\rm w})^3/3$ to the free energy in Eq.~\eqref{eq:f_bulk}. With a value $\beta\e_3=1.6$, we are able to obtain a phase diagram (not displayed) having better agreement with the experimental result. However, given that other cubic terms such as e.g.\ $-\e_3^{wwa}(n^{\rm w})^2n^{\rm a}$ must surely also be included if the term in $\e_3$ is also included, we prefer to not go down this route.

To improve the present theory, we would instead suggest going back to the Hamiltonian \eqref{eq:Hamiltonian} and seeking to incorporate aspects of the desired ordering at that level; for example, by replacing the Hamiltonian with one having terms like those in the Hamiltonian of Ref.~\cite{ackland2006structure}, which presents a lattice model for water-methanol mixtures incorporating orientational aspects of the molecular ordering. Then, a lattice DFT could be derived by building on the work of Ref.~\cite{zimmermann2023lattice}; see also Ref.~\cite{maeritz2021density}.

\subsection{Inhomogeneous Fluid}
The process of choosing parameters to match the experimental bulk fluid phase diagram just described leads to having the values of all the parameters in the model needed to solve the lattice DFT. However, because the DFT only needs dimensionless combinations of parameters, specifically the lattice spacing $\sigma$ (the length scale in the model) and the overall energy scale in the model $\beta$ don't have to be individually specified -- c.f.\ the dimensionless combinations in Eq.~\eqref{eq:epsilons}. This applies also when it comes to solve the model for inhomogeneous fluid situations, e.g.\ for calculating the density profiles corresponding to the free interface between coexisting phases. As discussed already in relation to the comparison presented in Fig.~\ref{fig:st}, we only need to give $\sigma$ a specific value if we want to compare with the experiments.

The surface tension results from our model presented in Fig.~\ref{fig:st} are calculated using the approach described in Ref.~\cite{hughes2014introduction}, here generalised to a ternary mixture. To calculate the surface tension and the corresponding equilibrium fluid interfacial density profiles, we assume a planar interface between the two coexisting phases that is perpendicular to one of the coordinate directions and then use Picard iteration to solve the Euler-Lagrange equations from minimising the grand potential
\begin{equation}
\Omega=F
-\sum_\ii\mu^{\rm w}n_\ii^{\rm w}
-\sum_\ii\mu^{\rm o}n_\ii^{\rm o}
-\sum_\ii\mu^{\rm a}n_\ii^{\rm a}.
\end{equation}
Thus, we solve the set of coupled equations
\begin{equation}
\frac{\partial \Omega}{\partial n_\ii^{p}}=0,
\end{equation}
for $p=\{{\rm a,o,w}\}$. This is done using periodic boundary conditions and, as mentioned already, we fix the oil chemical potential to be $\beta\mu_{\rm o}=-2$, so that the bulk system phase diagram is essentially the same as that of the incompressible mixture in Fig.~\ref{fig:phase_diag}. Calculating the free interface density profiles (and from these the corresponding surface tension values) for a range of different \red{values of the alcohol chemical potential $\mu_{\rm a}$, allows us to produce the full curve displayed in Fig.~\ref{fig:st}. This calculation is somewhat akin to the calculation of the binodal discussed in the previous section, which is the corresponding calculation for obtaining homogeneous bulk coexisting phases.}

\section{Concluding Remarks}
\label{sec:5}
We have presented an experimental determination of the phase diagram of an idealised Ouzo system comprising water, ethanol and trans-anethole. Our experiments are broadly in agreement with previous results for the Ouzo phase diagram \cite{Vitale2003, scholten2008life, tan2016evaporation, tan2018evaporation}. The additional benefit of the approach taken here is that we have been able to accurately determine the properties of the coexisting phases, including determination of tie-lines, measurement of sample densities and interfacial tensions and identified the plait point. 

In parallel, we have also developed a simple lattice DFT for both the bulk fluid phase behaviour and inhomogeneous fluid properties. The DFT captures the main features of the mixture thermodynamics and captures most of the key features of the phase diagram. It also faithfully reproduces the variations in the surface tension as the amount of alcohol in the system is varied.

The results presented here not only deepen our understanding of the bulk mixture phase behaviour, but by developing a theory for the inhomogeneous fluid, this paves the way for future studies of the nonequilibrium dynamics. This is because the free energy we have developed here can straightforwardly be input into dynamical density functional theory (DDFT) \cite{hansen2013theory, marconi1999dynamic, archer2006dynamical, te2020classical} and specifically the lattice-DDFT developed in Ref.~\onlinecite{chalmers2017dynamical}. This has already been done for the case of aqueous colloidal suspensions in contact with a range of different surfaces to describe the evaporative drying dynamics \cite{perez2021changing}.

An advantage of the modelling approach taken here, i.e.\ by developing a simple lattice DFT that captures the main features of the system, is that it can be adapted fairly simply to other multiphase mixtures. To apply our lattice DFT to other mixtures is largely a matter of choosing appropriate values of the interaction parameters $\e^{pq}$. There are previous studies showing how the bulk fluid phase diagram changes as these are varied \cite{davis1996statistical} and now by following the path laid out here, the homogeneous fluid properties and also dynamics can be investigated.

The behaviour of liquids in contact with surfaces is a particular important aspect worth investigating, due to its every-day relevance. As mentioned in the introduction, lattice DFTs of the type used here have already been used successfully in studies of the wetting behaviour of one-component liquids on surfaces \cite{hughes2014introduction, hughes2015liquid, buller2017nudged} adsorption studies \cite{kierlik2001capillary, woo2001mean, schneider2014filling}, binary liquid mixtures on surfaces \cite{woywod2003phase, WoSc2004jpm, robbins2011modelling, archer2010dynamical, chalmers2017modelling, areshi2024binding, perez2021changing} and various other applications \cite{edison2015critical, edison2015phase, tasios2016critical, archer2023stability}. The work presented here now paves the way e.g.\ to investigate the behaviour of ternary mixtures on planar surfaces \cite{mukherjee2022wetting}, or in any other situations where the fluid is in contact with surfaces.

\subsection{Implications for Ouzo emulsion stability}
\begin{figure}[b!] 
   \centering
   \includegraphics[width=1.0\linewidth]{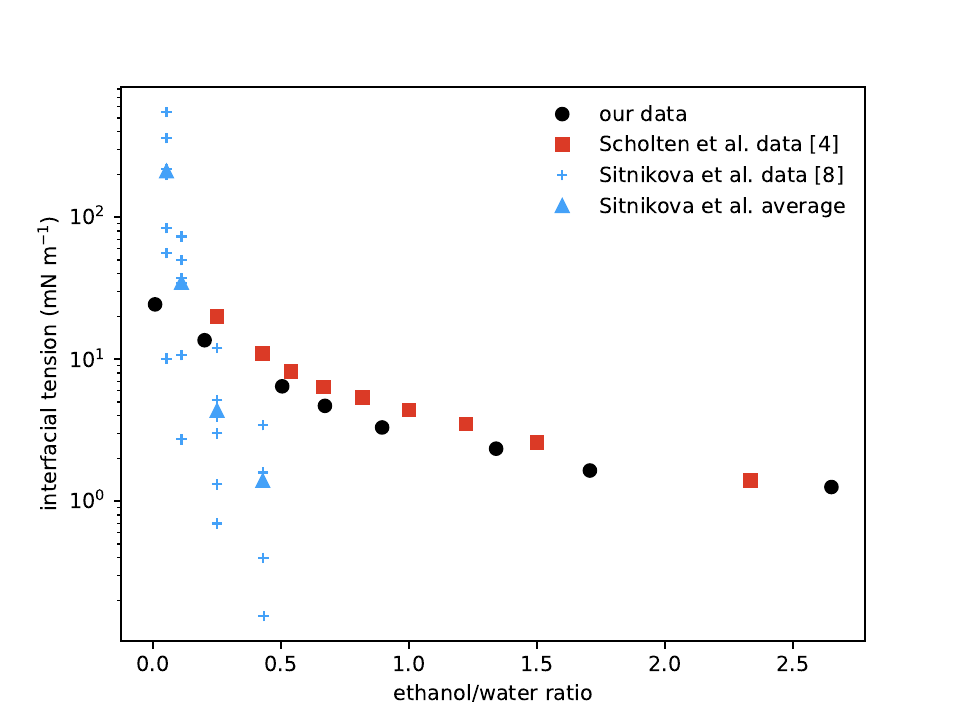} 
   \caption{Comparison of our surface tension measurements (black circles) with those of Scholten et al. \cite{scholten2008life} (red squares) showing good agreement. The blue data points are values deduced from the measured growth rates of spontaneous emulsions formed within the Ouzo region \cite{sitnikova2005spontaneously}, hence the narrow range of x-values. The vertical scatter in the individual measurements (small crosses) reflects variation with oil content, but averages (triangles) are the same order of magnitude as the values measured directly.}
   \label{fig:st_compare}
\end{figure}

A particular non-equilibrium aspect worth future investigation (and indeed one of our motivations for embarking on this study) is to understand the puzzling stability of the Ouzo effect, i.e.\ the long-lived emulsions that form when water is added to the Ouzo spirit \cite{scholten2008life}. The drink is typically diluted four parts to one. This corresponds to moving from the state point in the phase diagram where $(w_{\rm w},w_{\rm a},w_{\rm o})\approx(60,40,0.05)$ to the point $\approx(92, 8, 0.01)$, where the system demixes to form long-lived oil-rich droplets. The precipitated oil phase must have composition almost entirely depleted of water, with composition roughly (0,4,96), occupying a total sample volume of around 0.01\%. At this volume fraction, spacing between droplets would be almost 50 times their size, meaning collisions and coalescence between droplets would be unlikely. Any growth of oil droplets would be as molecular transport through the continuous water-rich phase, via the process of Ostwald ripening. 

\begin{figure}[t!] 
   \centering
   \includegraphics[width=1.0\linewidth]{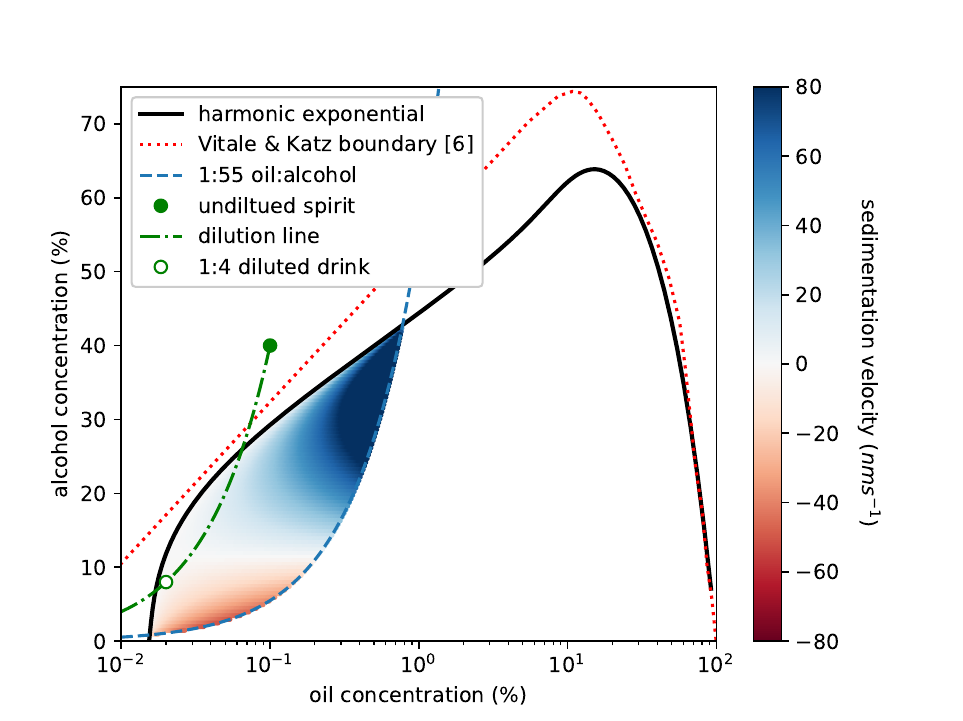} 
   \caption{Phase diagram of ethanol concentration against logarithm of oil concentration to expand the low oil corner. Colouring shows the Ouzo region where a stable emulsion has been previously reported\cite{Vitale2003}, located beneath the (solid black) cloud curve and above the (dashed blue) 1:55 oil to alcohol composition line. Blue shading represents sedimentation (positive speeds) and red indicates creaming (negative speeds) calculated for the spontaneously formed emulsion droplets, calculated using droplet radius data from Ref.\cite{Vitale2003} and density data from this study. There is a significant region (pale colouring) in which the droplets are close to being density matched ($\Delta \rho \approx 0$) and would sediment only a few mm in a week. The green circle marks the commercial Ouzo drink, the green dash-dot line shows the compositional change on diluting with water, and the open green circle is the serving suggestion, diluted with 4 parts water to 1 part spirit. This point is in the very slow sedimentation region. The red dashed line is the phase boundary from Ref.\cite{Vitale2003} for a system  using DVB in place of trans-anethole.}
   \label{fig:sed_compare}
\end{figure}

We can use information from our measurements to compare with Ostwald ripening rates as measured by Sitnikova et al.~\cite{sitnikova2005spontaneously}. 
By measuring the growth rates of oil droplets they were able to deduce a value for the interfacial tension of the droplets. 
In Fig.\ref{fig:st_compare} we compare their values to our directly-measured interfacial tensions. 
The scatter in their data for each value of ethanol/water ratio is due to different oil concentrations used, between 0.01\% and 0.6\%. Knowledge of the phase diagram suggests that such a small change in oil concentration is insufficient to significantly vary the coexisting compositions and so should not change the interfacial tension.
However, samples with more oil will contain more droplets, so their observed dependence could instead be due to increased coalescence or a higher concentration of dissolved molecular oil.
Notwithstanding this minor discrepancy, their inferred data is the same order of magnitude as our measurements, and also decreases with increasing ethanol. 
We also compare with direct measurements from Ref.~\cite{scholten2008life} which, although the phases are characterised slightly differently, are in good agreement.

We also use our measurements to reconsider sedimentation and creaming rates of the emulsion droplets. 
In Fig 4 of Ref.~\cite{Vitale2003}, Vitale and Katz show that droplet radius is predicted by oil (DVB not anethole) supersaturation and is independent of absolute ethanol concentration. 
We calculate a Stokes sedimentation velocity using a logarithmic fit to their data for droplet radius and a linear fit to our density data, assuming a constant viscosity for the continuous phase.
Fig.\ref{fig:sed_compare} plots this velocity throughout the Ouzo regime, bounded on the upper left side by the cloud curve. The right-hand side of the region is less clearly defined. For the DVB system Vitale and Katz suggested using the 1:20 oil to alcohol dilution line.
Our independent measurements show that for the anethole system, the Ouzo region terminates between the 1:50 and 1:60 dilution lines which is what is plotted as the green dot-dashed line.
We mirror their use of a logarithmic horizontal axis to magnify the low-oil region of the phase diagram. 
We predict speeds between 140 nm~s$^{-1}$ (1 cm per day sedimentation) to $-50$ nm~s$^{-1}$ (4mm per day creaming).
However for much of the region, in particular for drinks diluted with water from the Ouzo spirit, the speeds are around 100 times slower, which may well explain the long-term stability over weeks and even months.

With the DFT developed here, we are now ready as future work to consider an alternative explanation for the stability of such droplets, using an approach somewhat similar to that used recently to study the stability of airborne aerosol droplets \cite{archer2023stability}. We speculate that the small amount of alcohol in such droplets acts like a surfactant and stablises the water-oil interface. The present lattice DFT has been shown capable of describing such density enhancements at interfaces in other systems \cite{archer2023stability, areshi2024binding}, so we believe the present model will be useful to address such questions.

\section*{Acknowledgements}

A.J.A.\ and B.D.G.’s work on this was partially supported by the London Mathematical Society and the Loughborough University Institute of Advanced Studies.

D.J.F.\ and F.F.O.\ acknowledge the support of Nottingham Trent University summer bursary (PURS) scheme for funding Ross Broadhurst and James Rawlings.

\appendix
\section{Appendix}

The compositions of all cloud point samples plotted in Fig.~\ref{fig:binodal} are presented in Table~\ref{tab:binodal}. The properties of coexisting phases displayed in Fig.~\ref{fig:tielines} are listed in Table~\ref{tab:properties}, where we also list the measured interfacial tension values displayed in the plot in Fig.~\ref{fig:st}.

\begin{table*}[h!]
\small
    \caption{Compositions of cloud point samples, also displayed in Fig.~\ref{fig:binodal}.}
    \label{tab:binodal}
    \begin{tabular*}{\textwidth}{@{\extracolsep{\fill}}lll|lll|lll|lll|lll}
    \hline
    $w_{\rm w}$ & $w_{\rm o}$ & $w_{\rm a}$ & $w_{\rm w}$ & $w_{\rm o}$ & $w_{\rm a}$ & $w_{\rm w}$ & $w_{\rm o}$ & $w_{\rm a}$ & $w_{\rm w}$ & $w_{\rm o}$ & $w_{\rm a}$ & $w_{\rm w}$ & $w_{\rm o}$ & $w_{\rm a}$ \\
    \hline
    1.36	&	89.29	&	9.35	&	8.51	&	41.49	&	50.00	&	19.78	&	16.17	&	64.05	&	39.72	&	3.05	&	57.23	&	66.49	&	0.10	&	33.41	\\
    2.26	&	80.50	&	17.24	&	9.17	&	36.85	&	53.98	&	21.09	&	15.78	&	63.12	&	39.84	&	4.24	&	55.92	&	66.72	&	0.36	&	32.92	\\
    2.50	&	80.88	&	16.62	&	9.75	&	39.35	&	50.89	&	21.25	&	15.99	&	62.76	&	40.50	&	5.54	&	53.96	&	67.08	&	0.20	&	32.72	\\
    2.82	&	64.92	&	32.26	&	10.26	&	33.29	&	56.45	&	21.27	&	15.86	&	62.87	&	41.48	&	3.77	&	54.75	&	67.28	&	0.21	&	32.51	\\
    3.13	&	79.95	&	16.92	&	10.50	&	36.30	&	53.20	&	22.45	&	14.76	&	62.79	&	43.09	&	3.38	&	53.53	&	69.30	&	0.11	&	30.60	\\
    3.43	&	71.17	&	25.40	&	11.03	&	33.31	&	55.66	&	22.94	&	15.57	&	61.48	&	44.48	&	3.02	&	52.50	&	69.44	&	0.11	&	30.45	\\
    3.99	&	71.35	&	24.66	&	11.24	&	33.56	&	55.20	&	23.11	&	13.62	&	63.27	&	44.72	&	3.04	&	52.24	&	70.13	&	0.06	&	29.81	\\
    4.03	&	62.36	&	33.61	&	11.44	&	30.96	&	57.59	&	25.76	&	11.72	&	62.53	&	46.28	&	2.66	&	51.06	&	70.67	&	0.09	&	29.24	\\
    4.39	&	66.97	&	28.64	&	11.74	&	29.06	&	59.19	&	27.35	&	10.46	&	62.20	&	48.00	&	2.41	&	49.59	&	71.14	&	0.09	&	28.77	\\
    4.44	&	63.97	&	31.58	&	11.75	&	30.86	&	57.39	&	27.55	&	10.08	&	62.37	&	48.87	&	2.23	&	48.90	&	75.91	&	0.05	&	24.04	\\
    4.72	&	62.04	&	33.24	&	12.17	&	26.77	&	61.05	&	28.22	&	9.68	&	62.10	&	49.44	&	1.91	&	48.65	&	82.65	&	0.03	&	17.32	\\
    5.34	&	51.57	&	43.09	&	12.48	&	28.56	&	58.96	&	29.55	&	8.98	&	61.47	&	52.51	&	1.38	&	46.10	&	83.42	&	0.02	&	16.56	\\
    5.84	&	59.02	&	35.15	&	12.57	&	31.92	&	55.51	&	31.16	&	8.03	&	60.81	&	54.88	&	0.62	&	44.50	&	85.95	&	0.02	&	14.03	\\
    6.10	&	53.70	&	40.20	&	13.30	&	26.53	&	60.17	&	33.50	&	6.88	&	59.62	&	56.60	&	1.10	&	42.30	&	88.85	&	0.02	&	11.13	\\
    6.10	&	54.04	&	39.86	&	13.71	&	26.16	&	60.13	&	35.05	&	5.86	&	59.09	&	58.59	&	0.71	&	40.70	&	89.14	&	0.02	&	10.84	\\
    6.22	&	51.46	&	42.31	&	14.95	&	28.29	&	56.76	&	36.42	&	5.95	&	57.63	&	59.67	&	0.32	&	40.01	&	95.91	&	0.02	&	4.07	\\
    7.47	&	47.60	&	44.93	&	15.45	&	23.41	&	61.13	&	36.86	&	5.23	&	57.92	&	62.72	&	0.59	&	36.69	&	97.62	&	0.01	&	2.37	\\
    7.61	&	47.65	&	44.75	&	17.98	&	22.12	&	59.90	&	37.34	&	6.06	&	56.60	&	63.41	&	0.37	&	36.22	&	99.98	&	0.02	&	0.00	\\
    8.28	&	34.80	&	56.92	&	18.17	&	18.82	&	63.01	&	38.54	&	4.61	&	56.85	&	66.00	&	0.19	&	33.81	&		&		&		\\
    8.40	&	43.51	&	48.09	&	19.22	&	17.34	&	63.44	&	38.91	&	5.54	&	55.55	&	66.46	&	0.24	&	33.30	&		&		&		\\
    \hline
    \end{tabular*}
\end{table*}

\begin{table*}
\small
    \caption{Experimentally measured densities and interfacial tensions of coexisting phases, with mass fractions determined using the initial sample composition and applying the lever rule with knowledge of the fitted binodal and the origin point. It was not possible to measure the interfacial tension of all coexisting samples, and for one, there was such a small volume of the upper phase that its density could not be measured either. This data is plotted in Figs.~\ref{fig:tielines} and \ref{fig:st}.}
    \label{tab:properties}
    \red{
    \begin{tabular}{cccc|cccc|c}
    \hline
    \multicolumn{4}{c|}{\red{oil rich}}                & \multicolumn{4}{c|}{\red{water rich}}                  & {interfacial tension} \\
	water	&	oil	&	alcohol	&	density	&	water	&	oil	&	alcohol	&	density	&		\\	
	\%	&	\%	&	\%	&	g/cm$^3$	&	\%	&	\%	&	\%	&	g/cm$^3$	&	mN~m$^{-1}$	\\	\hline
0.15 & 99.85 & 0.00 & 0.986 & 99.98 & 0.02 & 0.00 & 0.997 & 24.35 \\
0.42 & 97.77 & 1.81 & 0.983 & 83.21 & 0.03 & 16.76 & 0.972 & 13.62 \\
0.59 & 96.04 & 3.37 & 0.981 & 75.84 & 0.05 & 24.11 & 0.957 &  \\
0.86 & 93.31 & 5.83 & 0.980 & 66.34 & 0.18 & 33.48 & 0.943 & 6.45 \\
1.10 & 90.89 & 8.01 & 0.978 & 59.52 & 0.50 & 39.97 & 0.930 & 4.70 \\
1.44 & 87.69 & 10.87 & 0.976 & 52.04 & 1.40 & 46.57 & 0.915 & 3.31 \\
2.12 & 81.51 & 16.37 & 0.974 & 40.86 & 4.37 & 54.77 & 0.894 & 2.35 \\
2.71 & 76.57 & 20.73 & 0.971 & 34.42 & 6.84 & 58.74 & 0.883 & 1.65 \\
3.78 & 68.31 & 27.91 & 0.971 & 26.82 & 10.62 & 62.56 & 0.875 &  \\
4.37 & 64.20 & 31.43 & 0.965 & 23.95 & 12.57 & 63.48 & 0.871 & 1.26 \\
5.13 & 59.32 & 35.55 & 0.967 & 21.06 & 15.08 & 63.85 & 0.871 &  \\
5.70 & 55.93 & 38.37 & 0.963 & 19.30 & 17.00 & 63.70 & 0.871 &  \\
6.89 & 49.51 & 43.59 &  & 16.38 & 21.08 & 62.54 & 0.887 &  \\
7.07 & 48.64 & 44.29 & 0.938 & 16.02 & 21.68 & 62.30 & 0.886 &  \\
7.26 & 47.71 & 45.03 & 0.927 & 15.65 & 22.34 & 62.02 & 0.883 &  \\ \hline
\end{tabular}}
\end{table*}


%

\end{document}